\PassOptionsToPackage{fontset=fandol}{ctex}
\documentclass{cpbtex}

\usepackage{algpseudocode}
\usepackage{graphicx}
\usepackage{dcolumn}
\usepackage[ruled]{algorithm}

\renewcommand{\today}{%
  \ifcase\month\or
    January\or February\or March\or April\or May\or June\or
    July\or August\or September\or October\or November\or December%
  \fi\space\number\day, \number\year}

\let\oldcite\cite
\renewcommand{\cite}[1]{\textsuperscript{\oldcite{#1}}}

\begin{document}

\title{PdrQC: Pauli-space Discriminative Representations based Quantum Classifier}

\author{
    Yuhang Tu(屠宇航)$^{1}$,
    Jinfan Wang(王劲凡)$^{1}$,
    Hao Huang(黄浩)$^{1}$,\\
    Le Wang(王乐)$^{1}$,
    Shengmei Zhao(赵生妹)$^{1}$, and
    Anqi Zhang(张安琪)$^{1}$%
    \thanks{Corresponding author.}\\
    $^{1}$Institute of Signal Processing and Transmission,\\
    Nanjing University of Posts and Telecommunications,\\
    Nanjing 210003, China
}

\date{\today}
\maketitle

\begin{abstract}
    Quantum classification faces two key challenges. First, the difficulty of distinguishing between different classes varies: some class pairs are easy to separate, while others are more challenging. Second, practical execution is affected by noise, finite sampling, and measurement overhead.
    To address these issues, we propose the Pauli-Space Discriminative-Representation based Quantum Classifier (PdrQC), a framework for task-adaptive multiclass quantum classification. The method evaluates candidate upload circuits using low-weight Pauli features and formulates upload design as a structured model selection problem based on discriminative representations.
    By progressively selecting upload structures and compact Pauli readout features for the target multiclass task, the framework achieves a better balance between classification accuracy and resource efficiency.
    {Numerical simulations} were conducted on the MNIST and Fashion-MNIST datasets with $K\in\{2,3,5,7,10\}$. The results demonstrate that PdrQC, through its task-adaptive Pauli representation, achieves an effective balance among multiclass classification accuracy, quantum-circuit complexity, and measurement overhead, making it suitable for multiclass quantum classification under limited hardware resources.

\end{abstract}


\section{Introduction}

Classification is a fundamental task in machine learning\cite{Cerezo2022Challenges}.
Before classification is performed, classical samples must be encoded into quantum states through a data-uploading circuit, which acts as a quantum feature transformation. A suitable upload structure can preserve and enhance class-discriminative information in the original data, whereas an unsuitable structure may suppress relevant differences and reduce classification accuracy. Circuit-centric and data-centric analyses have shown that the encoding map strongly affects the geometry of the induced representation space and the achievable decision boundaries\cite{Schuld2020CircuitCentric,Huang2021PowerData}. This observation forms the basis of feature-map- and kernel-based quantum learning\cite{Schuld2019FeatureHilbert,Havlicek2019Supervised}.

Existing quantum classifiers include multiclass kernels, variational
circuits, and hybrid quantum--classical models
\cite{Mernyei2023MulticlassKernel,Plekhanov2025Multiclass,Hou2023VQLP,Zhang2023SQC,Cheng2024HQCCNN,Wang2024HQDNN,Wu2025VQDA}.
Data re-uploading can increase expressivity \cite{PerezSalinas2020Data}, while
trainable kernels and gradient-free methods adapt quantum models to a task
\cite{Hubregtsen2022TrainingQEK,Glick2023KernelTraining,Mernyei2025KernelTraining,Zhang2024GCOPQC}.
However, these approaches commonly fix the upload structure
or optimize it for a kernel or variational objective, without using the final
measured multiclass representation itself to guide structure selection.

This limitation is further exacerbated under realistic execution conditions. Model performance is not determined solely by ideal-state separability, but is also affected by finite sampling, the selected readout observables, measurement grouping, and the available measurement budget \cite{Huggins2021Efficient,Crawford2021Efficient,Zhu2024Optimizing,Mao2024StateRecon}. An upload structure may generate a large pool of Pauli readout terms, but retaining all of them is neither necessary nor execution-efficient. Weakly discriminative or redundant Pauli terms increase the measurement cost, while noncommuting terms require additional measurement settings. Moreover, under a fixed shot budget, distributing shots over an excessive number of observables can increase estimation fluctuations and reduce feature stability.
Consequently, {the upload structure and its compact} Pauli readout
{should be selected together and evaluated under the same measurement budget}.

Motivated by these considerations, {we propose} the {Pauli-space
Discriminative Representations based Quantum Classifier} (PdrQC). PdrQC
{retains the Pauli-space-guided structured-selection principle:
candidate upload sequences are compared through the discriminative
representations that they generate in a common low-weight Pauli space.}
{It advances this principle to direct $K$-class classification by
selecting one compact Pauli coordinate set shared by all classes.} {The
same selected coordinates guide candidate evaluation and form the input to final
multinomial inference.}

The contributions are {threefold}. {First, we formulate upload design
as a progressive structured-selection problem in low-weight Pauli space:}
candidate {sequences} are evaluated {by their}
task-discriminative Pauli representations {without optimizing quantum-gate parameters.}
{Second, we develop a direct $K$-class Pauli representation with one
sparse coordinate set shared by all classes, used for both candidate evaluation
and final multinomial inference.}
{Third, we combine QWC grouping with fixed-budget finite-shot evaluation
to quantify the accuracy--resource trade-off and reduce readout and circuit costs.}

This work is a numerical simulation study. The remainder of the paper is organized as follows. Section~\ref{sec:PdrQC_method} {introduces data preprocessing and Pauli-space projection within the PdrQC method and then presents the complete classifier construction}. {Section~}\ref{sec:experiments} reports the simulation results, and Section~\ref{sec:conclusion} concludes the paper.

\section{PdrQC Method}
\label{sec:PdrQC_method}
\begin{figure*}[htbp]
    \centering
    \IfFileExists{figures/structurePdrQC.pdf}{%
        \includegraphics[width=0.98\linewidth]{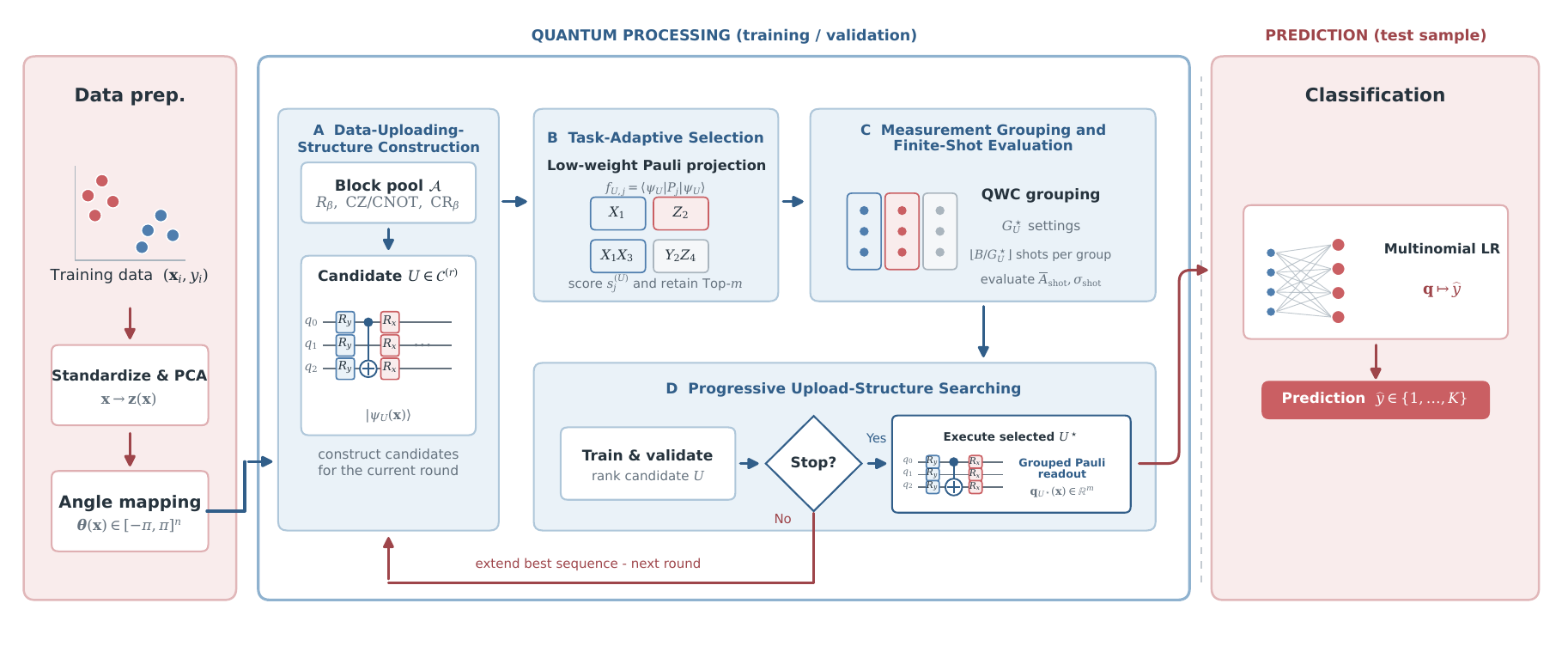}%
    }{%
        \fbox{\parbox[c][0.22\textheight][c]{0.94\linewidth}{%
                \centering Placeholder for the PdrQC framework diagram%
            }}%
    }
    \caption{Overall workflow of PdrQC. During model selection, standardized
        PCA features are angle-encoded by candidate upload structures. Each
        candidate is projected into a low-weight Pauli space, scored for
        multiclass discrimination, compressed into a qubit-wise commuting
        readout, and ranked using validation and finite-shot performance. The
        retained structure is progressively extended until the stopping
        condition is met. During inference, the selected upload circuit and
        grouped Pauli measurements generate a compact feature vector for direct
        multinomial classification.}
    \label{fig:structurePdrQC}
\end{figure*}
As shown in Fig.~\ref{fig:structurePdrQC}, the proposed PdrQC framework
constructs a task-specific multiclass classifier through a progressive search over
data-uploading structures and a task-adaptive selection of compact Pauli
readout features. {Standardization, PCA, and angle mapping provide
the classical preprocessing shown in the left panel of Fig.~\ref{fig:structurePdrQC}.}
{Candidate upload-structure construction corresponds to module~A.}
For each candidate upload structure, the encoded quantum
states are projected onto a low-weight Pauli feature space. The resulting
features are evaluated according to their direct multiclass classification
performance and Pauli-space separability. {Pauli projection and task-adaptive
selection correspond to module~B, measurement grouping and finite-shot
evaluation to module~C, and the progressive search to module~D.} The best candidate is progressively
extended for at most $R$ rounds. For the selected upload structure, a compact
Pauli subspace is retained, and a classical multinomial logistic classifier
directly learns the decision boundaries among all $K$ classes. {The selected
upload circuit is executed and the grouped Pauli readout is obtained within the
quantum-processing block, whereas multinomial classification and the final
prediction are shown in the right panel.}

{We now describe the quantum-processing steps in modules~A--D
of Fig.~\ref{fig:structurePdrQC}.}

\subsection{{Data-Uploading-Structure Construction}}
\label{sec:candidate_uploads}

{Consider a labeled data set
$\mathcal{D}=\{(\bm{x}_i,y_i)\}_{i=1}^{N}$, with
$\bm{x}_i\in\mathbb{R}^{d}$ and $y_i\in\{1,\ldots,K\}$.}

{PCA is used only to match the input dimension to the available
qubit count; task-specific discrimination is subsequently extracted from the
Pauli-space quantum representation.}

The input features are first standardized, and PCA is then fitted to the standardized training data. Let $z_k(\bm{x})$ denote the $k$-th retained principal component. It is mapped to a rotation angle according to
\begin{equation}
    \theta_k(\bm{x})
    =
    \operatorname{clip}\left(
    \alpha
    \frac{z_k(\bm{x})}{s_k},
    -\pi,
    \pi
    \right),
    \label{eq:angle_map}
\end{equation}
where
$s_k
    =
    \max_{\bm{x}\in\mathcal{D}_{\mathrm{tr}}}
    \left|
    z_k(\bm{x})
    \right|$,
and we set $\alpha=\pi/2$. If $p<n$, the remaining coordinates are padded with zeros.

{$\bm{\theta}(\bm{x})=[\theta_1(\bm{x}),\ldots,\theta_n(\bm{x})]^{\mathsf T}
\in[-\pi,\pi]^n$.}

The quantum block pool is defined as
\begin{equation}
    \mathcal{A}=
    \left\{
    R_x,
    R_y,
    R_z,
    \mathrm{CZ}_{\mathrm{ring}},
    \mathrm{CNOT}_{\mathrm{ring}},
    \mathrm{CR}_{x}^{\mathrm{ring}},
    \mathrm{CR}_{y}^{\mathrm{ring}},
    \mathrm{CR}_{z}^{\mathrm{ring}}
    \right\}.
    \label{eq:upload_library}
\end{equation}

{The $R_x$, $R_y$, and $R_z$ blocks apply data-dependent rotations to
each qubit, while CZ and CNOT apply controlled gates on the cyclic ring, where
$q^{+}$ denotes the next qubit.}

The controlled-rotation blocks also follow the ring topology. Consistent with the implementation, the angle on each ring connection is the average of the input angles associated with its control and target qubits:
\begin{equation}
    \mathrm{CR}_{\beta}^{\mathrm{ring}}
    \left(
    \bm{\theta}(\bm{x}_i)
    \right)=
    \prod_{q=1}^{n}
    \mathrm{CR}_{\beta}^{q,q^{+}}
    \left(
    \frac{
        \theta_q(\bm{x}_i)
        +
        \theta_{q^{+}}(\bm{x}_i)
    }{2}
    \right),
    \beta\in\{x,y,z\}.
    \label{eq:controlled_rotation_block}
\end{equation}

Thus, for sample $\bm{x}_i$, each controlled-rotation gate depends symmetrically on the two feature angles associated with its ring connection.

{An $L$-block upload sequence is written as
$U=[g_1,\ldots,g_L]$, with $g_\ell\in\mathcal A$; only rotation blocks depend
on $\bm{\theta}(\bm{x}_i)$.} Repeated application of these blocks produces the final
encoded quantum state
\begin{equation}
    \left|
    \psi_U(\bm{x}_i)
    \right\rangle=
    g_L
    \left(
    \bm{\theta}(\bm{x}_i)
    \right)
    \cdots
    g_2
    \left(
    \bm{\theta}(\bm{x}_i)
    \right)
    g_1
    \left(
    \bm{\theta}(\bm{x}_i)
    \right)
    |0\rangle^{\otimes n}.
    \label{eq:encoded_state}
\end{equation}

\subsection{Task-Adaptive Selection}
\label{sec:pauli_projection}

{For an $n$-qubit state, the Pauli strings
$\mathcal{P}_n=\{I,X,Y,Z\}^{\otimes n}$ form the operator basis, with weight
$w(P)=\sum_{q=1}^{n}\mathbf{1}[P_q\neq I]$. PdrQC uses the low-weight family
\begin{equation}
    \mathcal{P}_{n,\leq k}
    =
    \{P\in\mathcal{P}_n:\,1\leq w(P)\leq k\}
    =
    \{P_1,\ldots,P_M\},
    \qquad
    M=M_{n,\leq k}
    =
    \sum_{j=1}^{k}\binom{n}{j}3^j.
    \label{eq:pauli_enumeration}
\end{equation}
where $k$ is the maximum retained weight.}

{For candidate structure $U$, define}
\begin{equation}
    f_{U,j}(\bm{x}_i)
    =
    \left\langle
    \psi_U(\bm{x}_i)
    \right|
    P_j
    \left|
    \psi_U(\bm{x}_i)
    \right\rangle.
    \label{eq:pauli_expectation}
\end{equation}

{Collecting the $M$ coordinates gives
$\bm f_U(\bm x_i)=[f_{U,1}(\bm x_i),\ldots,f_{U,M}(\bm x_i)]^{\mathsf T}$.}

For each candidate upload structure $U$, PdrQC evaluates the discriminative contribution of every Pauli coordinate over the complete $K$-class task. All classes participate simultaneously in the evaluation. Therefore, the feature-selection procedure does not decompose the multiclass task into independent binary comparisons.

{Let $\mu_{c,j}^{(U)}$ and $\mu_j^{(U)}$ denote the class-specific and
global means, respectively.}

The between-class dispersion of Pauli coordinate $j$ is defined as
\begin{equation}
    B_j^{(U)}
    =
    \frac{1}{N_{\mathrm{tr}}}
    \sum_{c=1}^{K}
    N_c
    \left(
    \mu_{c,j}^{(U)}
    -
    \mu_j^{(U)}
    \right)^2.
    \label{eq:feature_between_class}
\end{equation}

This quantity measures the overall separation of the $K$ class means from the global mean. A large value of $B_j^{(U)}$ indicates that different classes produce substantially different responses on Pauli coordinate $j$.

The within-class dispersion is defined as
\begin{equation}
    W_j^{(U)}
    =
    \frac{1}{N_{\mathrm{tr}}}
    \sum_{c=1}^{K}
    \sum_{\left(\bm{x}_i,y_i\right)
        \in\mathcal{D}_{\mathrm{tr}}^{(c)}}
    \left(
    f_{U,j}(\bm{x}_i)
    -
    \mu_{c,j}^{(U)}
    \right)^2.
    \label{eq:feature_within_class}
\end{equation}

The quantity $W_j^{(U)}$ measures the overall variation of samples around their corresponding class means. A small value indicates that samples belonging to the same class produce relatively consistent responses on the corresponding Pauli coordinate.

Based on the two quantities, the direct multiclass discrimination score of Pauli coordinate $j$ is defined as
\begin{equation}
    s_j^{(U)}
    =
    \frac{
        B_j^{(U)}
    }{
        W_j^{(U)}
        +
        \varepsilon
    },
    \qquad
    \varepsilon>0,
    \label{eq:multiclass_pauli_score}
\end{equation}
where $\varepsilon$ is a small constant used to avoid numerical instability. A high value of $s_j^{(U)}$ indicates that the Pauli coordinate provides large separation among different classes while maintaining compact distributions within individual classes.

\subsection{Measurement Grouping and Finite-Shot Evaluation}
\label{sec:measurement_evaluation}

The candidate Pauli strings are ranked according to $s_j^{(U)}$. Under a prescribed feature budget $m_{\max}$, the compact Pauli subspace is selected as
\begin{equation}
    \mathcal{S}_U^{\star}=
    \operatorname{TopM}
    \left(
    \left\{
    s_j^{(U)}
    \right\}_{j=1}^{M},
    m_{\max}
    \right),
    \label{eq:selected_pauli_subspace}
\end{equation}
where $\operatorname{TopM}(\cdot,m_{\max})$ retains the $m_{\max}$ Pauli strings with the highest multiclass discrimination scores.

{The selected observables are greedily partitioned into QWC groups;
two strings are compatible if, on every qubit, their local operators are equal
or at least one is the identity. Let $G_U^{\star}$ be the number of groups.} Given a total shot budget $B$ per
sample, the number of shots allocated to each group is

\begin{equation}
    N_{\mathrm{shot}}
    =
    \left\lfloor
    \frac{B}{G_U^{\star}}
    \right\rfloor.
    \label{eq:shots_per_group}
\end{equation}

For a Pauli expectation value $a\in[-1,1]$, the variance of its finite-shot estimator is approximated by
\begin{equation}
    \operatorname{Var}
    \left(
    \widehat{a}
    \right)
    \approx
    \frac{
        1-a^2
    }{
        N_{\mathrm{shot}}
    }.
    \label{eq:shot_variance}
\end{equation}

{Finite-shot evaluation is repeated $H$ times; its mean accuracy,
standard deviation, and decrease from ideal accuracy are reported. The compact
feature vector $\bm q_U(\bm x)$ contains the selected Pauli expectations.}

The subset $\mathcal{S}_U^{\star}$ is shared by the complete $K$-class task. Measurement grouping and finite-shot evaluation quantify its inference cost and sampling stability, but do not participate in the Pauli-feature ranking process.

\subsection{{Progressive Upload-Structure Searching}}
\label{sec:progressive_search}

The data-uploading structure is determined through a greedy progressive search over atomic upload blocks.

{All atomic blocks are evaluated in the first round; thereafter, each
block is appended to the preceding winner, for at most $R$ rounds.}

For each candidate $U\in\mathcal{C}^{(r)}$, PdrQC generates the low-weight Pauli feature pool, calculates the task-adaptive discrimination scores, selects the compact Pauli subset $\mathcal{S}_{U}^{\star}$, and partitions the selected observables into $G_{U}^{\star}$ measurement groups. A multiclass classifier is then trained using the selected Pauli features.

{For candidate $U$, $A_{\mathrm{val}}(U)$ is its validation accuracy
under the compact Pauli representation.}

To jointly account for predictive performance, the accuracy and measurement cost are considered together. Therefore, the candidate-ranking tuple is defined as
\begin{equation}
    \mathcal{R}(U)=
    \left(
    A_{\mathrm{val}}(U),
    \overline{A}_{\mathrm{shot}}(U),
    \sigma_{\mathrm{shot}}(U)
    \right),
    \label{eq:candidate_ranking_tuple}
\end{equation}

The components of $\mathcal{R}(U)$ are compared lexicographically. Thus, ideal validation accuracy is assigned the highest priority, followed by finite-shot accuracy, and measurement cost.

{The search stops when validation accuracy reaches
$\tau_{\mathrm{acc}}$ or the depth reaches $R$.}

Because extending an upload sequence does not guarantee improved performance, the final upload structure is selected from the winners retained at all completed search rounds.

\section{{Numerical Simulations}}
\label{sec:experiments}

\subsection{{Simulation Setup}}
\label{subsec:experimental_setup}

{Numerical simulations} were conducted on the MNIST and Fashion-MNIST data sets, both of which contain ten classes. We considered classification tasks with \(K\in\{2,3,5,7,10\}\) classes. For each \(K<10\), ten tasks were randomly selected and evaluated, whereas the \(K=10\) setting corresponded to the unique ten-class task.

For an 8-qubit circuit, all Pauli observables with $w(P) \leq 2$ were included in the initial candidate pool. The pool therefore contained 276 observables, including single-qubit Pauli terms and two-qubit Pauli correlation terms. {To control readout complexity, PdrQC retained 28, 32, 40, 48, and 60 Pauli coordinates for $K=2,3,5,7,$ and $10$, respectively.}

The quantum-classifier comparison additionally included a data re-uploading classifier (DRC), following the layered variational data re-uploading principle of P{\'e}rez-Salinas \emph{et al.}~\cite{PerezSalinas2020Data}, and a quantum-kernel classifier (QKC), following the quantum feature-space kernel approach~\cite{Schuld2019FeatureHilbert,Havlicek2019Supervised}.

In the {numerical simulations}, the quantum encoding circuit was implemented with 8 qubits. To control the measurement complexity, only Pauli observables with weight up to 2 were considered, and between 28 and 60 Pauli coordinates were retained. The circuit architecture was determined by a progressive search strategy with at most 3 rounds. Finite-shot performance was averaged over 30 independent measurement realizations. The {simulation setup} is summarized in Table~\ref{tab:config}.

\begin{table}
    \centering
    \caption{\label{tab:config}
        {Simulation setup}.}
    \begin{tabular}{lc}
        \hline
        Quantity                                 & Value  \\
        \hline
        Number of qubits \(n\)                   & 8      \\
        Maximum Pauli weight \(k\)               & 2      \\
        Total Pauli features                     & 276    \\
        Selected Pauli features                  & 28--60 \\
        Target threshold \(\tau_{\mathrm{acc}}\) & 0.95   \\
        Total measurement shot budget \(B\)      & 2048   \\
        Finite-shot repetitions \(H\)            & 30     \\
        Maximum DRC training epochs              & 100    \\
        \hline
    \end{tabular}
\end{table}

\subsection{Contribution of the Pauli Feature Mapping}
\label{subsec:pca_ablation}
\begin{figure*}[htbp]
    \centering
    \includegraphics[width=0.98\linewidth]{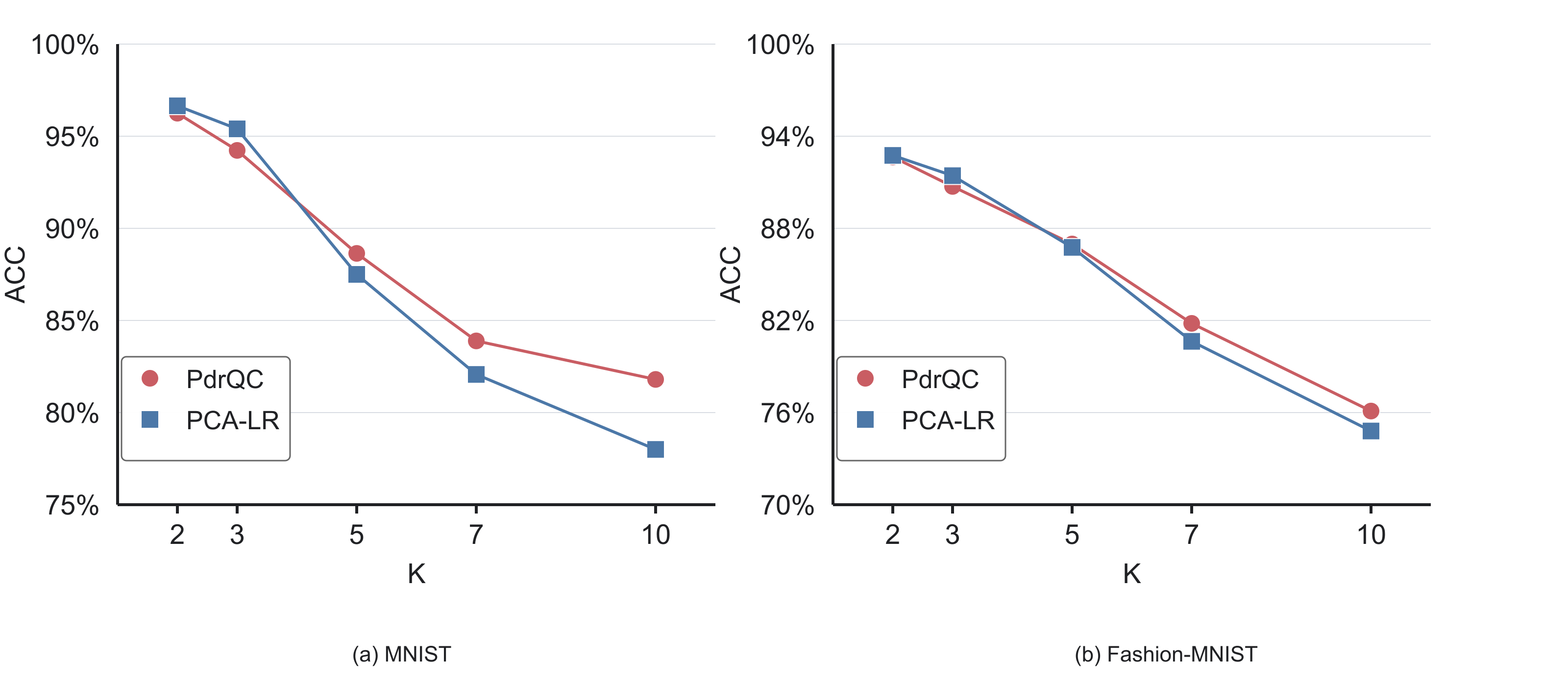}
    \caption{Accuracy of PdrQC and the PCA-LR on (a) MNIST and (b) Fashion-MNIST with different $K$. PCA-LR directly performs classification using the PCA features, whereas PdrQC further maps the PCA features into a quantum circuit and constructs a discriminative Pauli-selected feature subspace.}
    \label{fig:E1_PCA_ablation}
\end{figure*}
We removed the quantum module and directly trained a logistic regression (LR) classifier on the eight-dimensional PCA features. This PCA-LR baseline was used to determine whether the classification performance of PdrQC could be attributed primarily to PCA processing.

As shown in Fig.~\ref{fig:E1_PCA_ablation}, on MNIST, PdrQC achieved accuracies of 0.9625 and 0.9423 for $K=2$ and $K=3$, respectively, which were slightly lower than the corresponding PCA-LR accuracies of 0.9665 and 0.9540. However, when the number of classes increased to $K=5,7,$ and $10$, PdrQC achieved higher accuracies than PCA-LR. Similar performance was observed on Fashion-MNIST. These results show that PCA is sufficient to obtain competitive performance on tasks with relatively few classes. As the number of classes increases, however, the quantum Pauli-selected readout provides additional discriminative information that is not captured by the linear PCA representation.

\subsection{Effectiveness of the Pauli-Selected Readout}
\label{subsec:pauli_readout}
\begin{figure*}[htbp]
    \centering
    \includegraphics[width=0.98\linewidth]{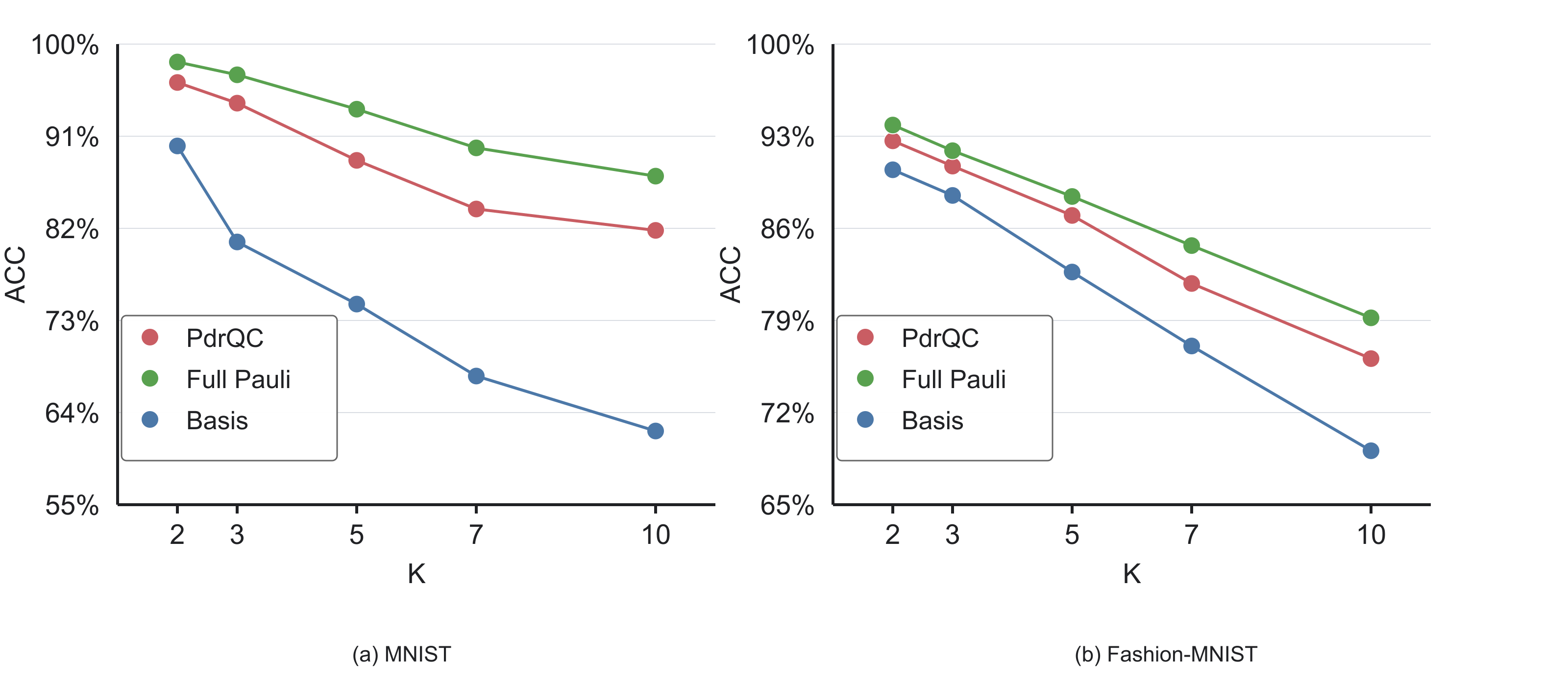}
    \caption{Accuracy obtained using different readout spaces on (a) MNIST and (b) Fashion-MNIST. PdrQC uses a compact selected Pauli subspace, Full Pauli uses all 276 Pauli observables with weight no greater than two, and computational-basis readout (Basis) uses the 256 output probabilities obtained by measuring the 8-qubit quantum state in the computational basis.}
    \label{fig:E2_readout_ablation}
\end{figure*}
We compared three readout mechanisms while keeping the input encoding circuit unchanged.

As shown in Fig.~\ref{fig:E2_readout_ablation}, Full Pauli achieved the highest accuracy in all evaluated tasks, indicating that the complete low-weight Pauli space preserves the largest amount of quantum-state information. Nevertheless, PdrQC retained accuracy close to that of Full Pauli while using substantially fewer observables. For example, at $K=10$, PdrQC used only 60 of the 276 candidate observables, corresponding to a $78.3\%$ reduction in the Pauli feature space. Under this compression, the accuracy decreased from 0.8710 to 0.8180 on MNIST and from 0.7920 to 0.7610 on Fashion-MNIST, respectively.

Both Pauli readouts substantially outperformed the computational-basis readout. These results indicate that the classification information is not fully represented by computational-basis probabilities. Overall, Full Pauli provides the highest classification accuracy, whereas PdrQC achieves a more favorable trade-off between predictive performance and the size of the measured Pauli space.

\subsection{Pauli-Space and Measurement Resource Efficiency}
\label{subsec:resource_efficiency}
\begin{figure*}[htbp]
    \centering
    \includegraphics[width=0.98\linewidth]{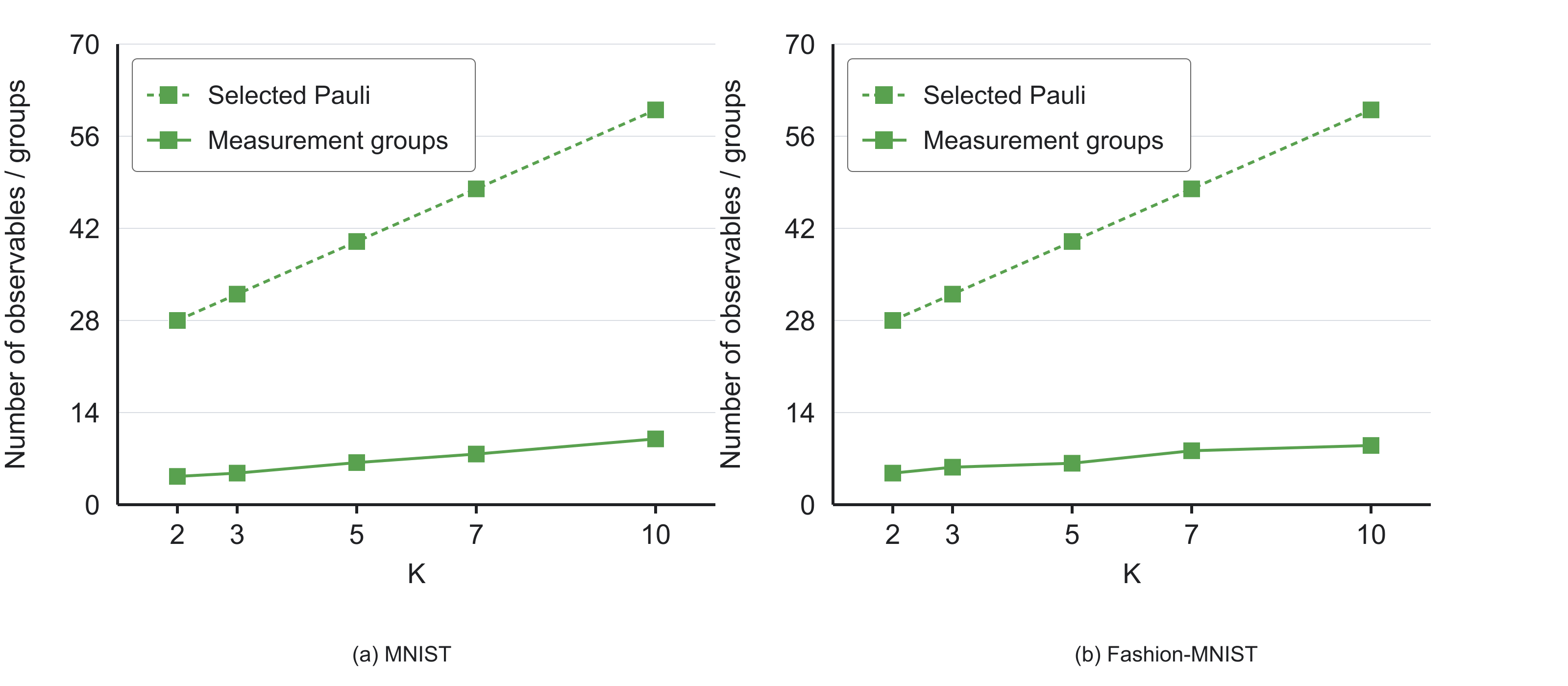}
    \caption{Scaling of PdrQC measurement resources with the number of classes $K$ on (a) MNIST and (b) Fashion-MNIST, where the dashed lines denote the number of selected Pauli observables and the solid lines denote the number of measurable groups.}
    \label{fig:E5_resource_scaling}
\end{figure*}
Fig.~\ref{fig:E5_resource_scaling} illustrates the numbers of selected Pauli observables and jointly measurable groups used by PdrQC. As the number of classes increased from $K=2$ to $K=10$, the number of selected Pauli observables increased from 28 to 60. These selected observables accounted for only $10.1\%$--$21.7\%$ of the complete 276-dimensional Pauli candidate pool, corresponding to feature-space reductions of $78.3\%$--$89.9\%$. The number of measurement groups increased from 4.3 to 10.0 on MNIST and from 4.8 to 9.0 on Fashion-MNIST. Therefore, although PdrQC retained between 28 and 60 Pauli observables, measurement grouping reduced their implementation to approximately 4 to 10 distinct measurement settings.

These results demonstrate that PdrQC constructs a compact task-dependent subspace from the complete Pauli candidate pool. By selecting Pauli observables according to their discriminative scores and subsequently exploiting measurement grouping, the method limits both the number of measured observables and the number of distinct quantum measurement settings.

\subsection{Performance Comparison with Quantum Classifiers}
\label{subsec:finite_shot}
\begin{figure*}[htbp]
    \centering
    \includegraphics[width=0.98\linewidth]{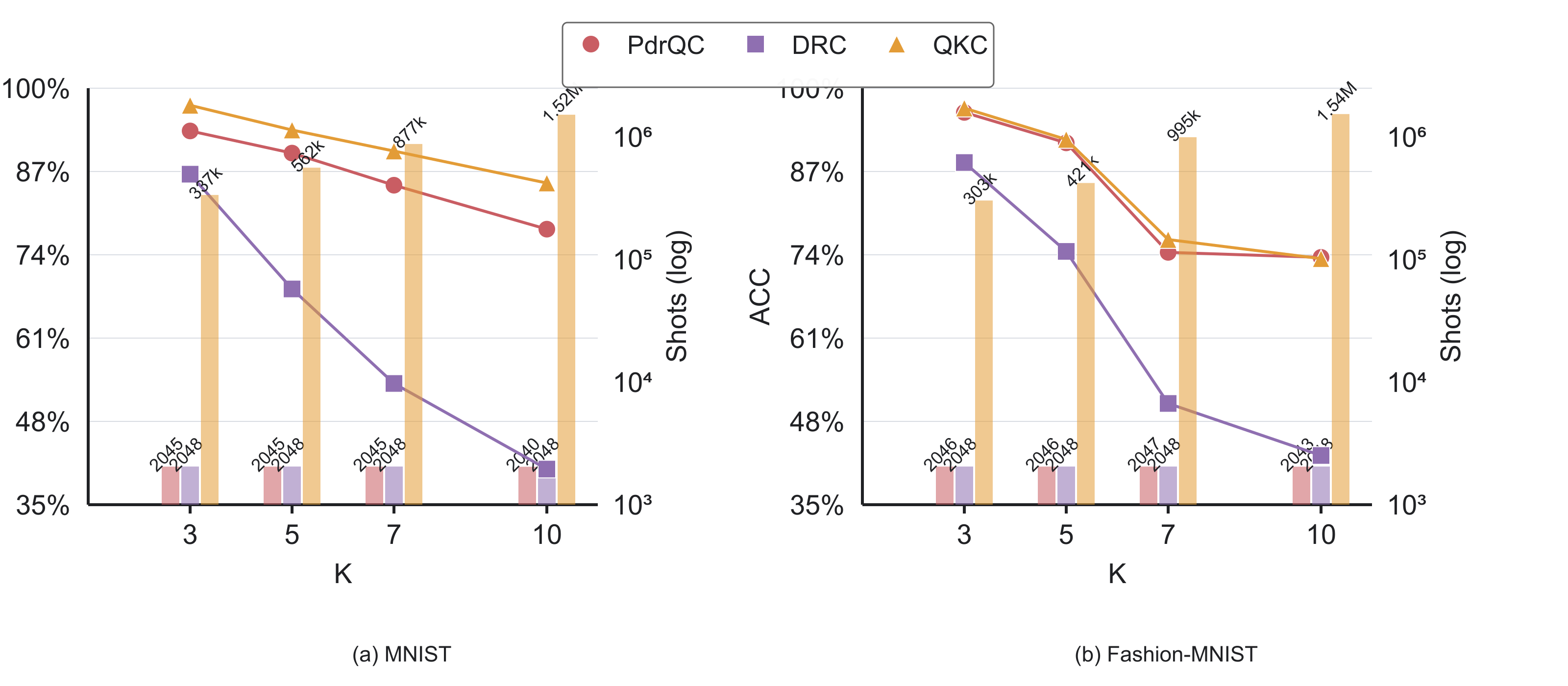}
    \caption{Accuracy (left axis) and measurement cost (right axis) of PdrQC, the data re-uploading classifier (DRC), and the quantum-kernel classifier (QKC) on (a) MNIST and (b) Fashion-MNIST.}
    \label{fig:E6_finite_shots}
\end{figure*}
We further compared the three quantum classifiers under finite-shot measurement. Fig.~\ref{fig:E6_finite_shots} illustrates that, for PdrQC, a total shot budget of $B=2048$ was distributed among the selected measurement groups. The finite-shot accuracy was averaged over 30 independent measurement realizations using the same frozen model.
This finite-shot surrogate models sampling uncertainty only; it does not include gate noise, decoherence, crosstalk, or biased readout errors.

Under the finite-shot setting, PdrQC outperformed the DRC on both data sets. On MNIST, the accuracy advantage increased from 6.75 percentage points at $K=3$ to 37.44 percentage points at $K=10$. On Fashion-MNIST, the corresponding advantage increased from 7.82 to 30.92 percentage points. Thus, PdrQC retained substantially stronger multiclass performance under a measurement budget comparable to that of the data re-uploading classifier.

The quantum-kernel classifier generally achieved higher accuracy, particularly on MNIST, but required a substantially larger number of circuit measurements. Across the two data sets, its measurement cost ranged from 303,104 to 1,538,048 shots per test sample, whereas PdrQC required only approximately 2,048 shots. The resulting quantum-kernel measurement cost was approximately 148--753 times that of PdrQC. In addition, at $K=10$ on Fashion-MNIST, PdrQC slightly outperformed the quantum-kernel classifier despite using far fewer shots.

These results show that the higher accuracy of the quantum-kernel classifier is obtained at a substantially increased inference cost. PdrQC instead provides a more compact balance between multiclass accuracy and test-time measurement resources, while maintaining nearly constant shot consumption as the number of classes increases.

\begin{figure*}[htbp]
    \centering
    \includegraphics[width=0.98\linewidth]{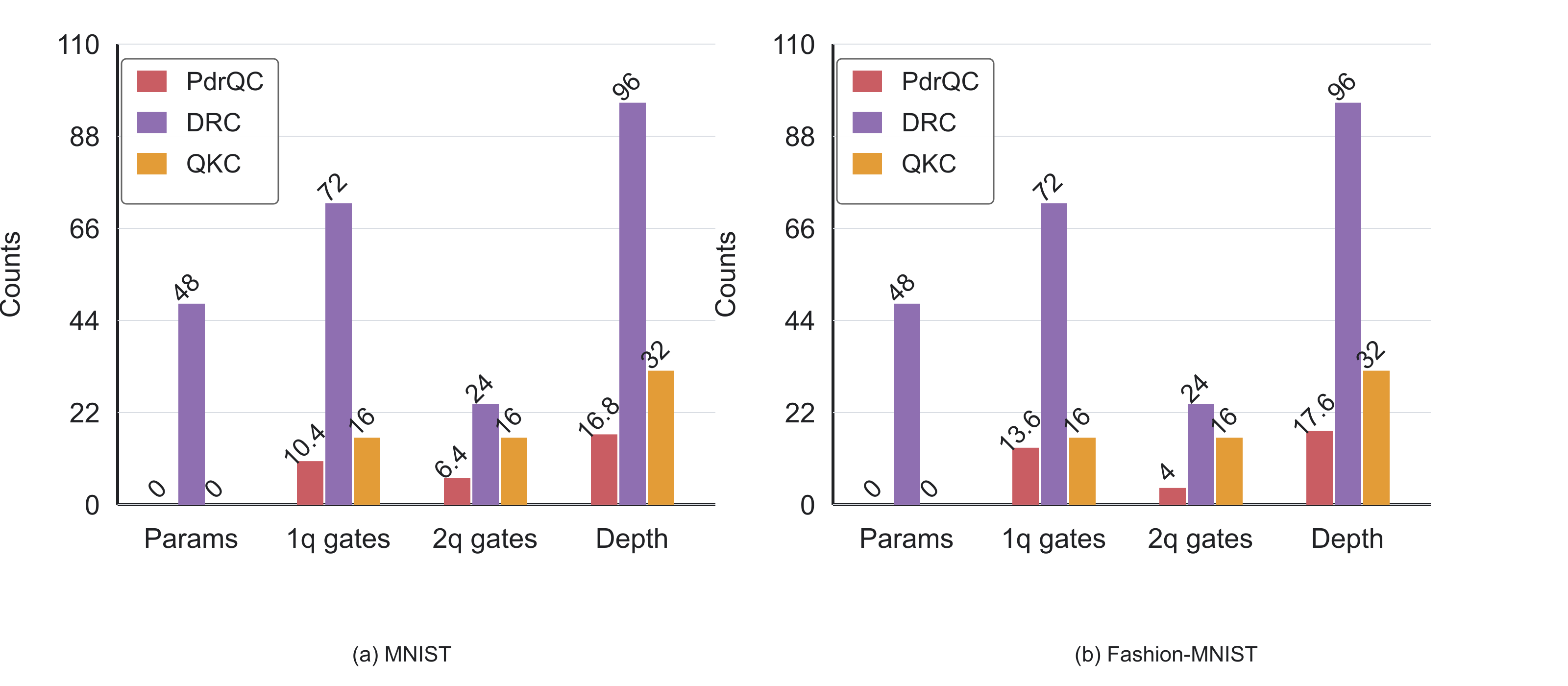}
    \caption{Quantum-circuit resource counts for PdrQC, the data re-uploading classifier (DRC), and the quantum-kernel classifier (QKC) on (a) MNIST and (b) Fashion-MNIST.}
    \label{fig:E4_circuit_cost}
\end{figure*}

We further compared the quantum-circuit configurations of the three quantum classifiers in terms of the number of trainable quantum parameters (Params), single-qubit gates (1Q gates), two-qubit gates (2Q gates), and circuit depth.

Figure~\ref{fig:E4_circuit_cost} shows that PdrQC requires substantially fewer quantum-circuit resources than DRC and QKC. Unlike DRC, which contains 48 trainable quantum parameters, PdrQC does not require quantum-parameter optimization. On MNIST, PdrQC uses 10.4 single-qubit gates, 6.4 two-qubit gates, and a circuit depth of 16.8, reducing the depth by $82.5\%$ relative to DRC and by $47.5\%$ relative to QKC. On Fashion-MNIST, PdrQC uses 13.6 single-qubit gates, 4.0 two-qubit gates, and a circuit depth of 17.6, corresponding to depth reductions of $81.7\%$ and $45.0\%$ relative to DRC and QKC, respectively. The substantially smaller number of two-qubit gates indicates lower circuit-execution complexity. Although this reduction may be beneficial for future noisy-hardware execution, the present {numerical simulations} do not model hardware noise. Overall, PdrQC achieves a more compact circuit implementation without introducing trainable quantum parameters.

\clearpage
\section{Conclusion}
\label{sec:conclusion}
This work developed PdrQC, a multiclass classification framework that formulates upload design as a structured selection problem based on separability, measurement cost, and finite-sampling effects.
{Numerical simulations} on MNIST and Fashion-MNIST with $K\in\{2,3,5,7,10\}$ show that the selected Pauli features provide additional discriminative information beyond that provided by PCA features and computational-basis readout.
Under finite-shot evaluation, PdrQC consistently outperforms the data re-uploading classifier and achieves performance comparable to that of the quantum-kernel classifier on Fashion-MNIST while requiring substantially fewer shots.
It also uses fewer quantum gates and lower circuit depth, while requiring no independently trainable quantum parameters.
These results demonstrate that task-adaptive Pauli representations enable an effective balance among multiclass accuracy, quantum-circuit complexity, and measurement overhead, making PdrQC suitable for quantum image classification under limited hardware resources.

Future work will extend the method to hardware noise, larger data sets, and more flexible upload-structure and Pauli-feature search strategies.

\section*{Data availability statement}
The data that support the findings of this study are openly available. 

\section*{Acknowledgment}
This work was supported by the National Natural Science Foundation of China under Grant No. 62375140.

\bibliographystyle{iopart-num}
\bibliography{references}

\end{document}